\documentclass[
preprint,showpacs,
amsmath,amssymb,aps
]{revtex4-1}

\usepackage{graphicx}% Include figure files
\usepackage{dcolumn}% Align table columns on decimal point
\usepackage{bm}% bold math
\newcommand{\md}{\mathrm d}

\begin{document}
%\preprint{APS/123}

\title{Gauge dependence on the chiral phase transition of QCD\\
at finite temperature in the Schwinger--Dyson equation}

\author{Hiroaki Kohyama}
%\email{author@institution.edu}
\affiliation{Department of Physics,
National Taiwan University, Taipei 10617, Taiwan.
}

\date{\today}

%%%%%%%%%%%%%%%%%%%%%%%%%%%%%%
\begin{abstract}
We study the gauge dependence on the chiral phase transition of
Quantum chromodynamics at finite temperature based on the
quenched Schwinger--Dyson equation. We first solve the equations
without approximations at finite temperature in general gauge,
then study the gauge dependence on the critical temperature of
the chiral phase transition. We find that the critical temperature
drastically depends on the choice of the gauge, and the parameters
at the quenched level in the Schwinger--Dyson equations.
\end{abstract}
%%%%%%%%%%%%%%%%%%%%%%%%%%%%%%

% 11.15.-q, 
% PACS, the Physics and Astronomy Classification Scheme.
% \keywords{Suggested keywords}
% Use showkeys class option if keyword display desired

\maketitle
%\tableofcontents

%%%%%%%%%%%%%%%%%%%%%%%%%%%%%%
%%%%%%%%%%%%%%%%%%%%%%%%%%%%%%
\section{\label{sec:intro}
Introduction}
%%%%%%%%%%%%%%%%%%%%%%%%%%%%%%
%%%%%%%%%%%%%%%%%%%%%%%%%%%%%%
The chiral symmetry in quark matters is broken due to the nonperturbative
effect of the strong interaction. While it is expected to be restored at high
temperature since the nature of asymptotic freedom makes the strong
coupling weak. Therefore, it is natural to expect that there must be some
critical temperature on the chiral phase transition. The issue is interesting,
and it has been studied for decades both from theoretical and experimental
sides.

The Schwinger--Dyson equation (SDE) ~\cite{Dyson:1949ha} is
suitable approach for the investigation of the chiral phase transition,
and it is actually often employed in this context (for reviews, 
see,~\cite{Roberts:1994dr, Roberts:2000aa, Maris:2003vk,
Fischer:2006ub}). The solutions for the
equations tell us the field strength renormalization and the dynamically
generated mass, then we can directly see whether the chiral symmetry
is broken from the obtained solutions. It is known that even for 
quantum electrodynamics (QED) the chiral symmetry can be broken
when the coupling is strong enough, and there has been various works
on the SDE in QED (see, e.g.,~\cite{%
%%%%%%
Johnson:1964da,
Maskawa:1974vs,
FK76,
Ball:1980ay,
Roberts:1989mj,
Curtis:1990zs,
Burden:1993gy,
Bando:1993qy,
Fukazawa:1999aj,
Kizilersu:2009kg,
Bashir:2011dp,
Kizilersu:2014ela}).
%%%%%%
The coupling is strong enough to cause the symmetry breaking in
quantum chromodynamics (QCD), then the investigation becomes 
important and there also has been a lot of analyses based on the
SDE approach~\cite{%
Akiba:1987dr,
Kalashnikov:1987td,
Barducci:1989eu,
Blaschke:1997bj,
Maris:1997hd,
Harada:1998zq,
Ikeda:2001vc,
Harada:2001tr,
Takagi:2002vj,
Fueki:2002mm,
Krassnigg:2003dr,
Krassnigg:2006ps,
Harada:2007gg,
Blank:2010bz,
Mueller:2010ah,
Nakkagawa:2011ci,
Mader:2011zf%
}.
%%%%%%

The SDE is derived from quantum field theory, and the physical
predictions should be gauge, regularization and parameter independent
in principle. However, it is impossible to determine the exact form of
the equations which include infinite tower of terms. Therefore, for
the sake of the manageability of the equations, we inevitably apply
approximations, which causes the gauge dependence of the system.
The gauge dependence can clearly be seen in the quenched form,
where the gauge parameter explicitly appears in the equations.
It is, therefore, important to investigate the gauge dependence in
the practical analyses, since the physical predictions are expected
to crucially depend on the
gauge as observed in the strong coupling QED~\cite{Kizilersu:2014ela}.
Then, in this paper, we are going to perform the analyses in general
gauge and study the gauge dependence on the solutions, particularly
its influence on the critical behavior of the chiral phase transition.

The plan of the paper is following: Section~\ref{sec:SDE} presents the
SDE in QCD, the parameters in the equations and the effective potential
of the system based on the Cornwall-Jackiw-Tomboulis formalism.
We then perform the actual numerical analyses on the solutions for
SDE and the chiral phase transition in Sec.~\ref{sec:nr}. The concluding
remarks are given in Sec.~\ref{sec:conclusion}.

%%%%%%%%%%%%%%%%%%%%%%%%%%%%%%
%%%%%%%%%%%%%%%%%%%%%%%%%%%%%%
\section{\label{sec:SDE}
Schwinger--Dyson equation}
%%%%%%%%%%%%%%%%%%%%%%%%%%%%%%
%%%%%%%%%%%%%%%%%%%%%%%%%%%%%%
We present the procedure on how we study the phase transition
of the chiral symmetry breaking based on the SDE approach in
this section. We first show the SDE and perform parameter
fitting, then discuss the phase transition by using the effective
potential derived from the solutions of the equations.

%%%%%%%%%%%%%%%%%%%%%%%%%%%%%%
\subsection{\label{subsec:equation}
The equation}
%%%%%%%%%%%%%%%%%%%%%%%%%%%%%%
The equation for the quark self-energy $\Sigma(p)$ is given by 
%%%%%%
\begin{equation}
  \Sigma (p)
  = i g^2 \int \frac{{\mathrm d}^4 q}{(2\pi)^4}
     \gamma^{\mu} \frac{\lambda^a}{2}
     D_{\mu\nu}(p-q) S(q) \Gamma_{a}^{\nu}(p,q) ,
\end{equation}
%%%%%%
with  the strong coupling, $g$, the Gell-Mann matrices 
in the color space, $\lambda_a$, the propagators for the gluon
and quark, $D_{\mu \nu}(p-q)$ and $S(q)$, and the vertex
function $\Gamma_a^\nu (p,q)$. For $D_{\mu \nu}$ and
$\Gamma_a^\nu$, we use the following quenched forms
%%%%%%
\begin{eqnarray}
  &&D_{\mu\nu}(k)
  =  \frac{-g_{\mu \nu}}{k^2}
    +(1-\xi) \frac{k_{\mu}k_{\nu}}{k^4},
  \label{eq:photon_prop} \\
  &&\Gamma_a^{\nu}(p,q) = t_a \gamma^{\nu},
\end{eqnarray}
%%%%%%
where $\xi$ is the gauge parameter, $k_\mu = p_\mu -q_\mu$
and $t^a = \lambda^a/2$.

The general form of the quark propagator is given by
%%%%%%
\begin{equation}
  S(p_0, p) = \frac{1}
  { C(p_0,p)\gamma_0 p_0
    +A(p_0,p)\gamma_i p^i -B(p_0,p)}
\end{equation}
%%%%%%
with $p = |{\bf p}|$. Note that, contrary to the four dimensional form
used in the zero temperature case, the time and space components
are treated separately at finite temperature since the time component
becomes discrete due to the applied boundary condition. In more
concrete, the continuous $q_0$ integral is replaced by the discretized
summation 
%%%%%%
\begin{equation}
  \frac{1}{2\pi i} \int_{-\infty}^{\infty} \md q_{0} F(q_0)
  \to
  T \sum_{m=-\infty}^{\infty} F(i \omega_m),
\end{equation}
%%%%%%
here the frequency,  $\omega_m$, runs as
$\omega_m = (2m + 1)\pi T$ for the quark field~\cite{Bellac:2011kqa}.

Performing a bit of algebras we arrive at the following equations
%%%%%%
\begin{align}
 &C(\omega_n, p) = 1
   + C_2 \frac{\alpha_s}{\pi}
  %\int_{-\infty}^{\infty} \md q_{0}
  T \sum_{m=-\infty}^{\infty}
  \int_{\delta}^{\Lambda} \md q \,
     \left[ 
        {\mathcal I}_{CC} C(\omega_m, q)
     +{\mathcal I}_{CA} A(\omega_m, q)
     \right]      \Delta(\omega_m, q) , 
  \label{eq:C} \\
&A(\omega_n, p) = 1
   + C_2 \frac{\alpha_s}{\pi}
  %\int_{-\infty}^{\infty} \md q_{0}
  T \sum_{m=-\infty}^{\infty}
  \int_{\delta}^{\Lambda} \md q \,
     \left[
        {\mathcal I}_{AC} C(\omega_m, q)
    + {\mathcal I}_{AA} A(\omega_m, q)
     \right]      \Delta(\omega_m, q), 
  \label{eq:A} \\
&B(\omega_n, p) = m_0
   +  C_2 \frac{\alpha_s}{\pi}
  %\int_{-\infty}^{\infty} \md q_{0}
  T \sum_{m=-\infty}^{\infty}
  \int_{\delta}^{\Lambda} \md q \,
     \left[ {\mathcal I}_{B} B(\omega_m, q)
     \right]      \Delta(\omega_m, q),
  \label{eq:B}
\end{align}
%%%%%%
where $C_2=(N_c^2-1)/(2N_c)$ is the quadratic Casimir operator for
the color $SU(N_c)$ group,  $\alpha_s = g^2/(4\pi)$, $m_0$ is the
current quark mass, $\Delta(\omega_m, q)$ is given by
%%%%%%
\begin{align}
  \Delta(\omega_m, q)
  =  \frac{1}{ C^2(\omega_m, q) \omega_m^2  
                 + A^2(\omega_m, q) q^2 
                 + B^2(\omega_m, q)},
\end{align}
%%%%%%
and
%%%%%%
\begin{align}
  &  {\mathcal I}_{CC} = 
     -\xi_{+} \frac{{\omega_m}}{\omega_n}  I_1
     +2 \xi_{-} \frac{{\omega_m}}{\omega_n} {\omega^{\prime}_m}^2 I_2, \\
  & {\mathcal I}_{CA} = 
       \xi_{-} \frac{\omega^{\prime}_m}{\omega_n}  I_1
     +\xi_{-} \frac{\omega^{\prime}_m}{\omega_n}  
        \left[ {\omega^{\prime}_m}^2 - q^2 + p^2 \right] I_2, \\
  & {\mathcal I}_{AC} = 
      -\xi_{-} \frac{{\omega_m} {\omega^{\prime}_m}}{p^2}  I_1
      -\xi_{-} \frac{{\omega_m} {\omega^{\prime}_m}}{p^2}  
        \left[ {\omega^{\prime}_m}^2 + q^2 - p^2 \right] I_2, 
  \label{eq:AC}      \\
  & {\mathcal I}_{AA} = 
      -\frac{2q^2}{p^2}  
      -\frac{1}{2p^2} 
      \left[ \xi^{-}_3 {\omega^{\prime}_m}^2 
             + \xi_+(q^2+p^2)
      \right] I_1
      -\frac{1}{2p^2} \xi_{-} 
        \left[ {\omega^{\prime}_m}^4 - (q^2-p^2)^2 \right] I_2,
   \label{eq:AA}       \\
   &  {\mathcal I}_{B} = \xi^{+}_3 I_1 , \\
   &I_1 = \frac{q}{2p}
         \ln \frac{{\omega^{\prime}_m}^2+(q-p)^2}
                       {{\omega^{\prime}_m}^2+(q+p)^2}, \\
  &I_2 = \frac{q}{2p}
        \left[
            \frac{1}{{\omega^{\prime}_m}^2+(q-p)^2}
          -\frac{1}{{\omega^{\prime}_m}^2+(q+p)^2}
        \right].
\end{align}
%%%%%%
Here ${\omega^{\prime}_m} \equiv {\omega_n}-{\omega_m}$,
$\xi_{\pm} \equiv 1 \pm \xi$ and $\xi^{\pm}_3 = 3 \pm \xi$.
We introduced the infrared and ultraviolet cutoffs, $\delta$ and
$\Lambda$,  since the integral diverges for ultraviolet region and
the equations are not well defined in the $p^2 \to 0$ limit as
seen from the explicit forms for ${\mathcal{I}_{AC}}$ and
${\mathcal{I}_{AA}}$ in Eqs.~(\ref{eq:AC}) and (\ref{eq:AA}). 
The equations~(\ref{eq:C}), (\ref{eq:A}) and (\ref{eq:B}) are the
set of the SDE in QCD which we are going to solve in this paper.
In the following, we drop the suffix $s$ in $\alpha_s$ and
write the coupling strength as $\alpha$ for notational simplicity.

%%%%%%%%%%%%%%%%%%%%%%%%%%%%%%
\subsection{\label{subsec:parameter}
Parameters}
%%%%%%%%%%%%%%%%%%%%%%%%%%%%%%
The massless equations have three parameters, the strong coupling
$\alpha$, the three-momentum cutoff $\Lambda$ and the gauge
parameter $\xi$. Note that in the actual numerical calculations we
need to set the maximum number of the Matsubara frequency
$N_{t}$, and the infrared cutoff $\delta$. However, the results are
not affected if one chooses large and small enough values for $N_t$
and $\delta$, so we fix these values for $N_t = 50$ and
$\delta=0.01\Lambda$, in which we numerically confirmed that
these are indeed large and small enough for our analyses.

We first fix the cutoff parameter $\Lambda$ by using the empirical
value for the dynamical mass, $M(\omega_n,p) \equiv
B(\omega_n,p)/A(\omega_n,p)$, at $(\omega_n,p)=(\omega_0,\delta)$
to be $M = 335$MeV for various $\alpha$ in the Landau gauge.
Table~\ref{tab:para} aligns the values of $\Lambda$ for several
$\alpha$,
%%%%%%%%%%%%%%%%%%%%%%%%%%
\begin{table}[h!]
\caption{Parameters.}
\label{tab:para}
%\begin{ruledtabular}
\begin{center}
\begin{tabular*}{4cm}{@{\extracolsep{\fill}}cr}
\hline
$\alpha$  & $\Lambda$\hspace{0.7cm} \\
\hline
$1.5$ & $1582$MeV \\
$2.0$ &  $764$MeV   \\
$2.5$ &  $518$MeV   \\
$3.0$ &  $402$MeV   \\
\hline
\end{tabular*}
\end{center}
%\end{ruledtabular}
\end{table}
%%%%%%%%%%%%%%%%%%%%%%%%%%
where one sees that the coupling strength becomes weak when the
cutoff is large. The tendency is consistent with the expectation by
the renormalization group analysis. With these fitted parameters,
we study the gauge dependence on the chiral phase transition in
the next section.

%%%%%%%%%%%%%%%%%%%%%%%%%%%%%%
\subsection{\label{subsec:potential}
Effective potential}
%%%%%%%%%%%%%%%%%%%%%%%%%%%%%%
Before proceeding the numerical analyses, we remark the point
that the solutions derived from the equation give the stational
condition, not the global minimum of the potential. Therefore,
when one considers the phase transition, the search based on the
effective potential is important.

Below shows the form of the effective potential calculated from the 
Cornwall-Jackiw-Tomboulis (CJT) formalism~\cite{Cornwall:1974vz},
%%%%%%
\begin{align}
  {\mathcal V}_{\rm CJT}
  =  &\int \!\! \frac{\md^4 q}{i(2\pi)^4} {\rm tr}
  \Bigl[ \ln [S(q)] - S_0^{-1} (q) S(q)
  \Bigr] \nonumber \\
  & + C_2 \frac{g^2}{2}
     \int \!\! \frac{\md^4 q}{i(2\pi)^4}
     \int \!\! \frac{\md^4 k}{i(2\pi)^4}
     {\rm tr} \left[ S(q)\gamma_\mu S(k) \gamma_\nu \right]
     D^{\mu \nu} (q-k),
\end{align}
%%%%%%
with $S_0^{-1}(q) = q_\mu \gamma^\mu$.
The difference between the broken phase (Nambu--Goldstone phase)
and symmetric phase (Wigner phase),
%%%%%%
\begin{align}
  {\mathcal V}_{\rm rel}
  &=  {\mathcal V}[S_{\rm NG}] -  {\mathcal V}[S_{\rm W}] \nonumber \\
  &=  \frac{1}{\pi^2} T \sum_n
  \int_{\delta}^{\Lambda} \md q q^2
  \biggl[
     \ln \left(\frac{\Delta_{\rm NG}}{\Delta_{\rm W}} \right) 
     - \frac{C_{\rm NG}(\omega_n,q) \omega_n^2
     + A_{\rm NG}(\omega_n,q) q^2 }
     {C_{\rm NG}^2(\omega_n,q) \omega_n^2 
     + A_{\rm NG}^2(\omega_n,q) q^2 + B_{\rm NG}^2(\omega_n,q)}
     \nonumber \\
  & \qquad \qquad 
     + \frac{C_{\rm W}(\omega_n,q)  \omega_n^2
     + A_{\rm W}(\omega_n,q) q^2}
     {C_{\rm W}^2(\omega_n,q) \omega_n^2 
     + A_{\rm W}^2(\omega_n,q) q^2}   
  \biggr],
\end{align}
%%%%%%
is crucial to see which phase is energetically favored, then we study
the above potential difference for the determination of the phase
as done in~\cite{Harada:2007gg}.

%%%%%%%%%%%%%%%%%%%%%%%%%%%%%%
%%%%%%%%%%%%%%%%%%%%%%%%%%%%%%
\section{\label{sec:nr}
Numerical results}
%%%%%%%%%%%%%%%%%%%%%%%%%%%%%%
%%%%%%%%%%%%%%%%%%%%%%%%%%%%%%
Having presented the equations, the parameters and the effective
potential, we think it is ready for carrying on the actual numerical
calculations. With the help of the iteration method, we obtain the
solutions of the SDE, then analyze their gauge dependence and the
critical phenomena on the chiral phase transition in this section.

%%%%%%%%%%%%%%%%%%%%%%%%%%%%%%
\subsection{\label{subsec:ns}
Numerical solutions}
\begin{figure}[!h]
\begin{center}
  \includegraphics[width=5.4cm,keepaspectratio]{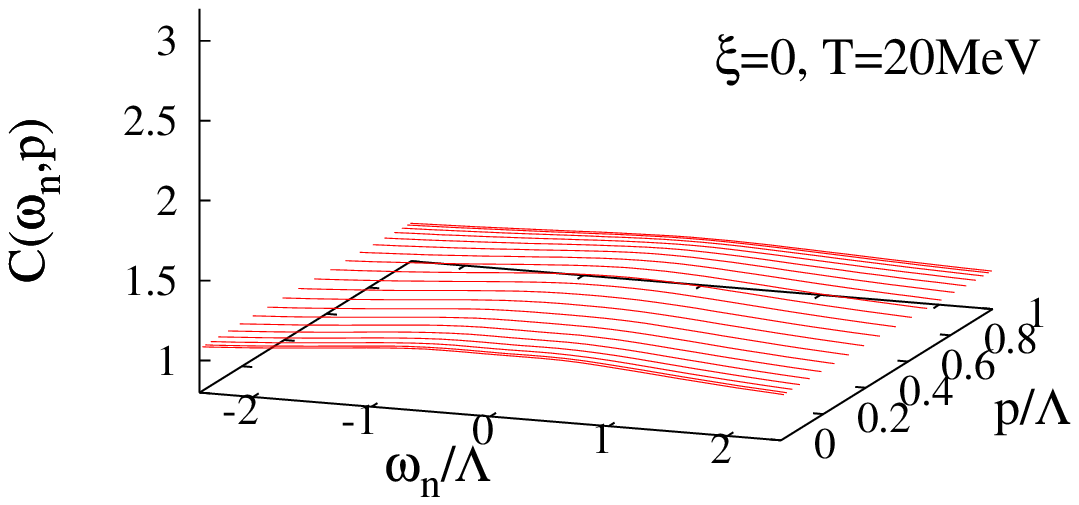}
  \includegraphics[width=5.4cm,keepaspectratio]{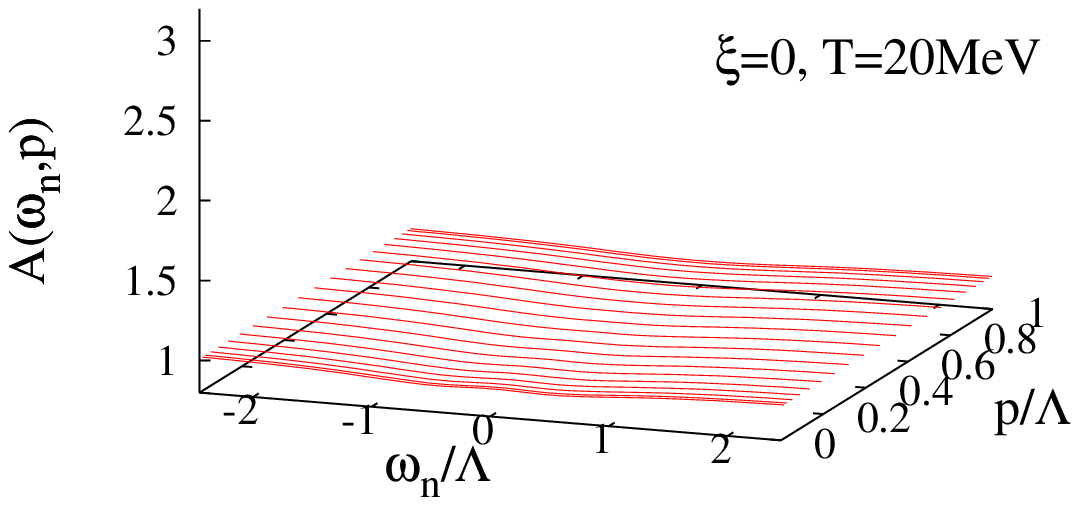}
  \includegraphics[width=5.4cm,keepaspectratio]{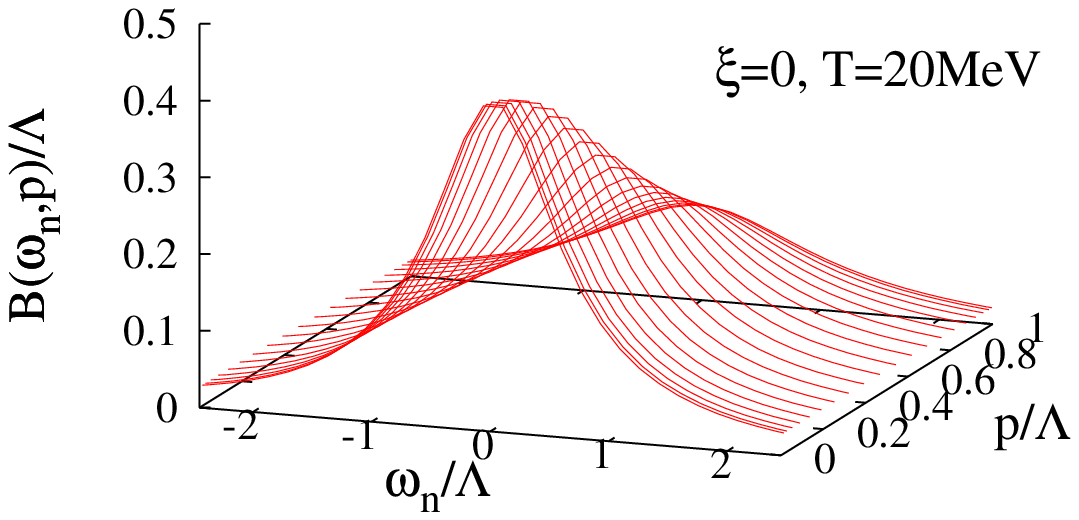}
  \includegraphics[width=5.4cm,keepaspectratio]{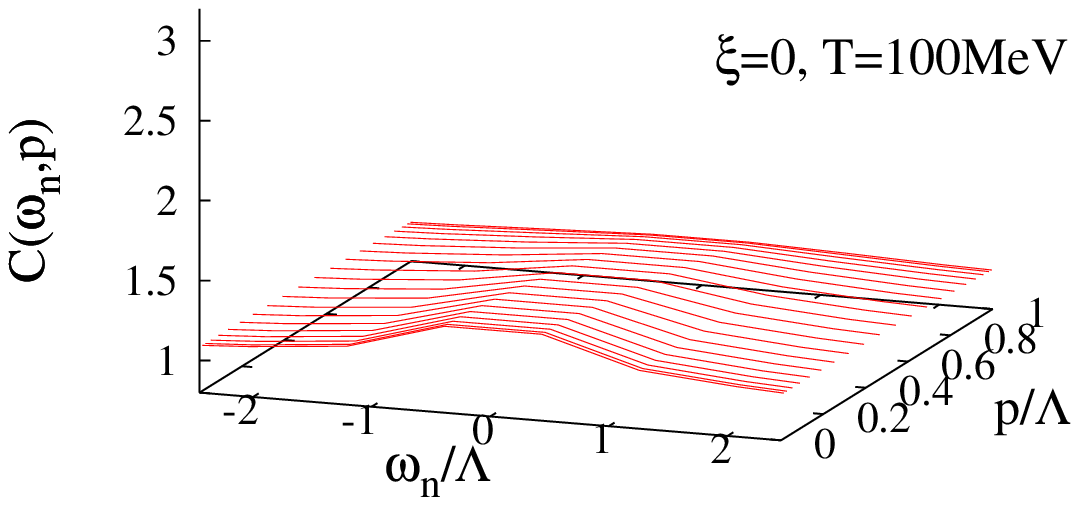}
  \includegraphics[width=5.4cm,keepaspectratio]{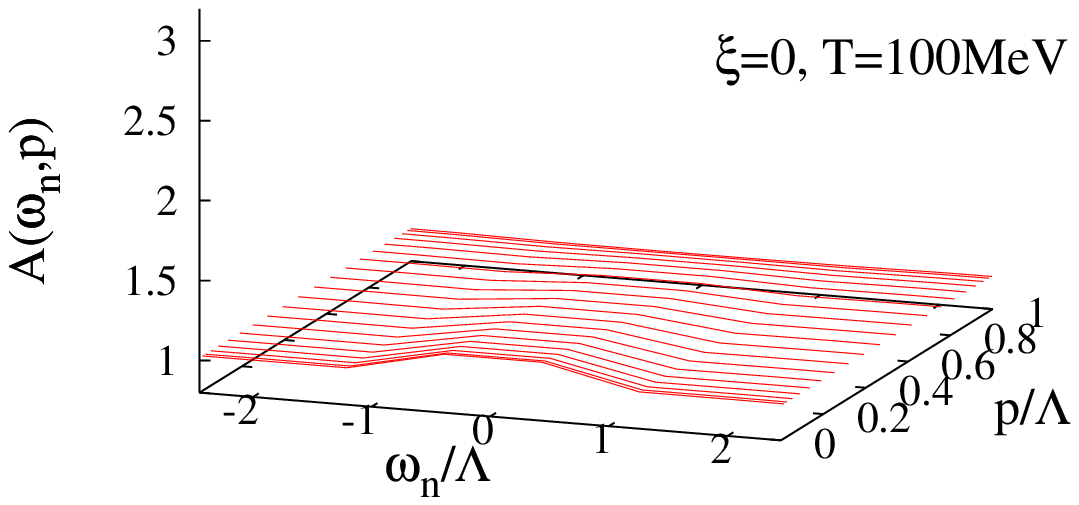}
  \includegraphics[width=5.4cm,keepaspectratio]{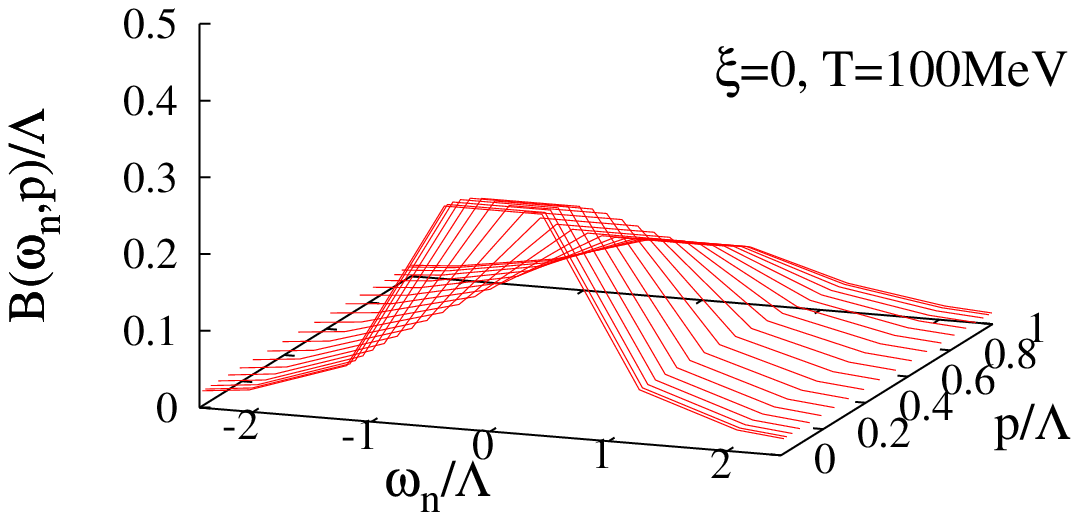}  
  \includegraphics[width=5.4cm,keepaspectratio]{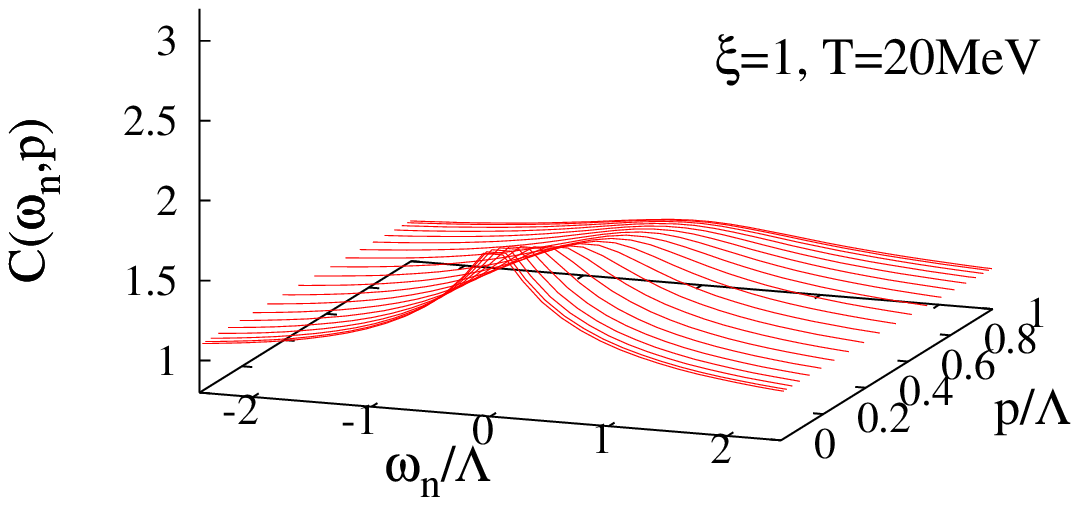}
  \includegraphics[width=5.4cm,keepaspectratio]{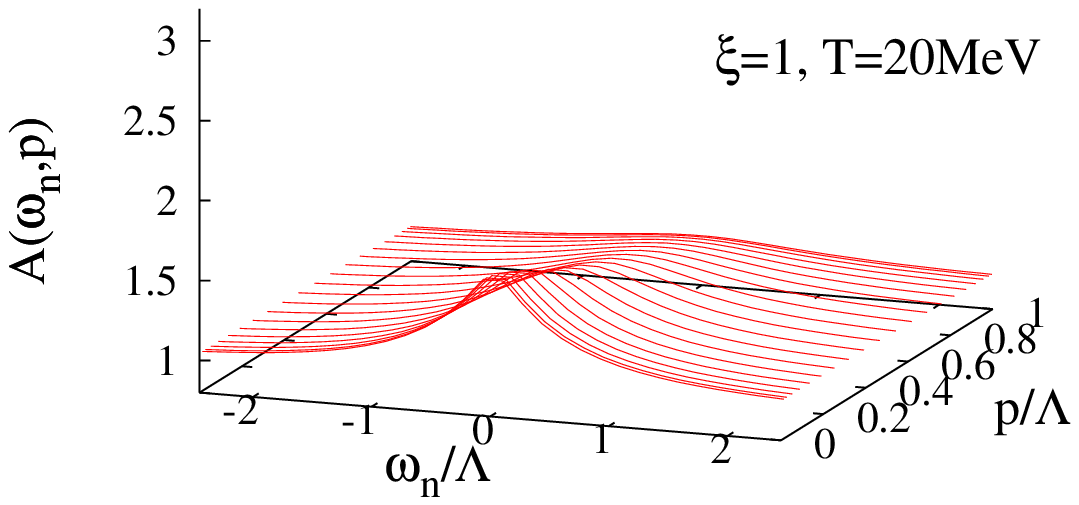}
  \includegraphics[width=5.4cm,keepaspectratio]{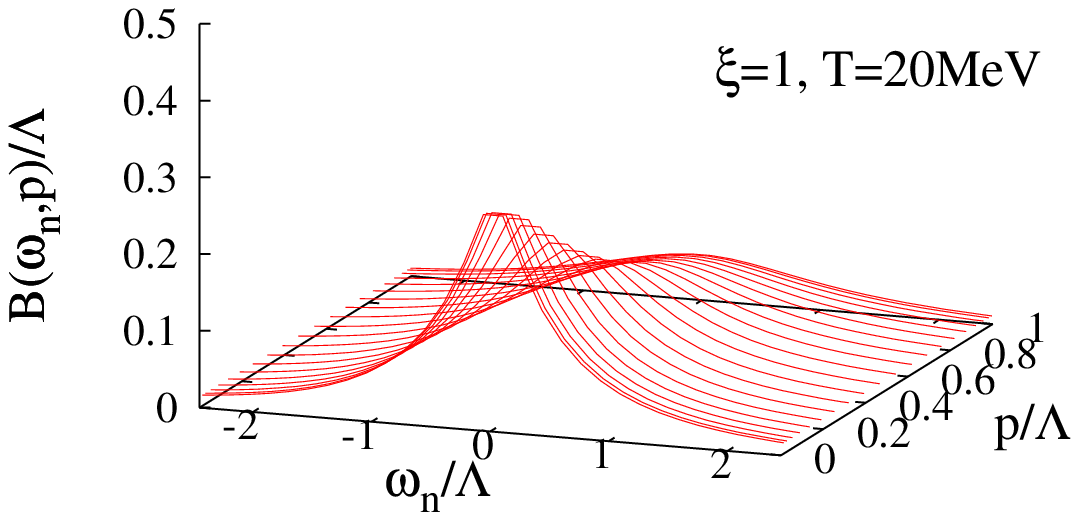}
  \includegraphics[width=5.4cm,keepaspectratio]{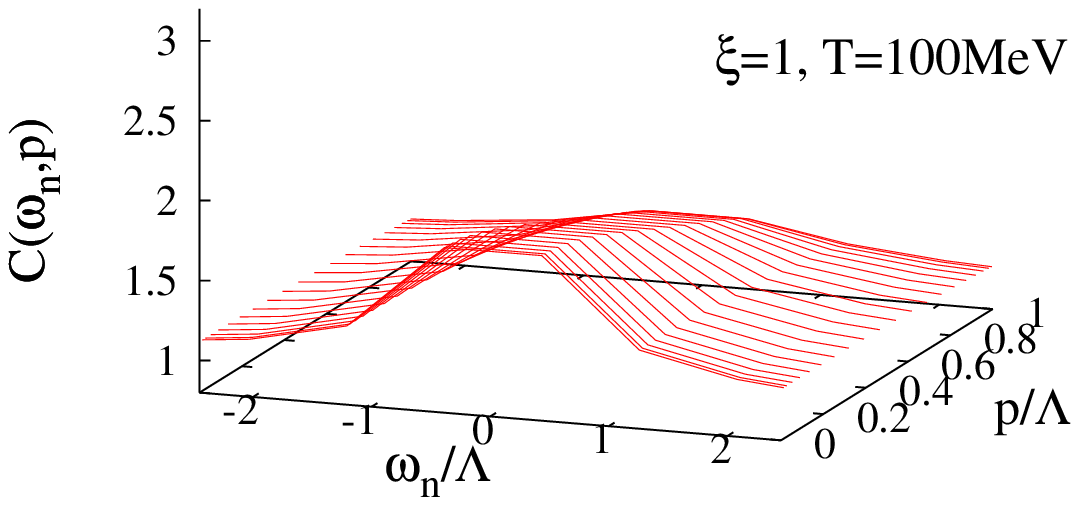}
  \includegraphics[width=5.4cm,keepaspectratio]{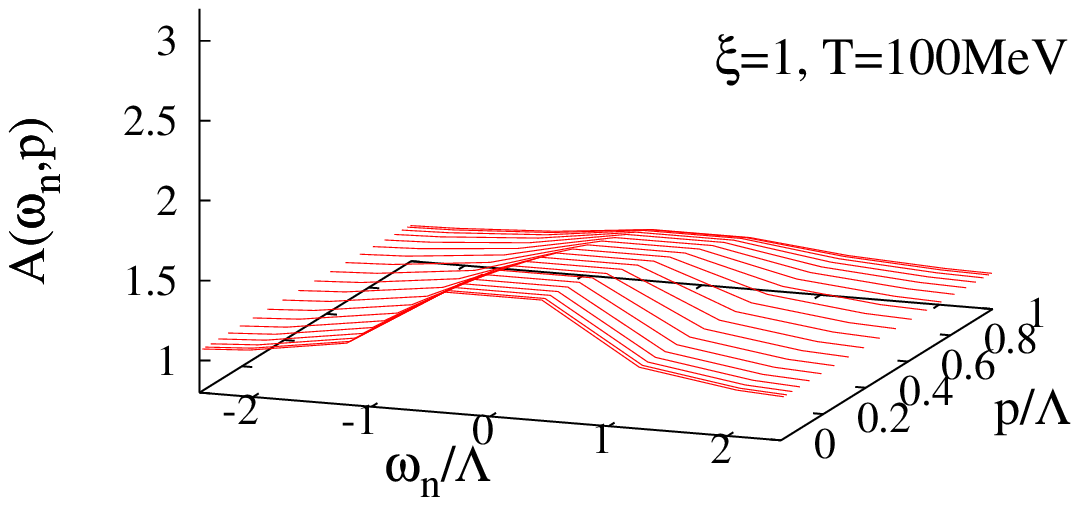}
  \includegraphics[width=5.4cm,keepaspectratio]{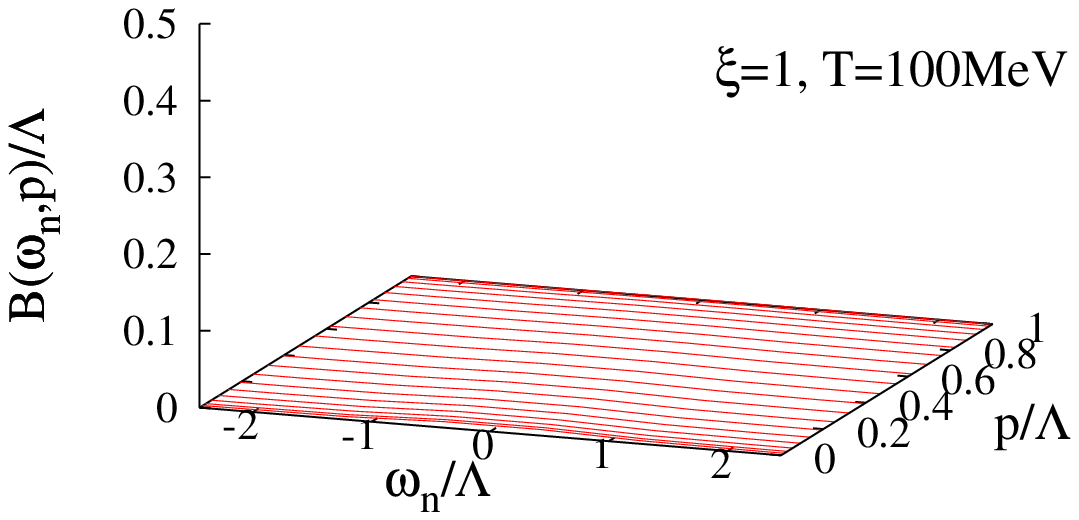}  
  \includegraphics[width=5.4cm,keepaspectratio]{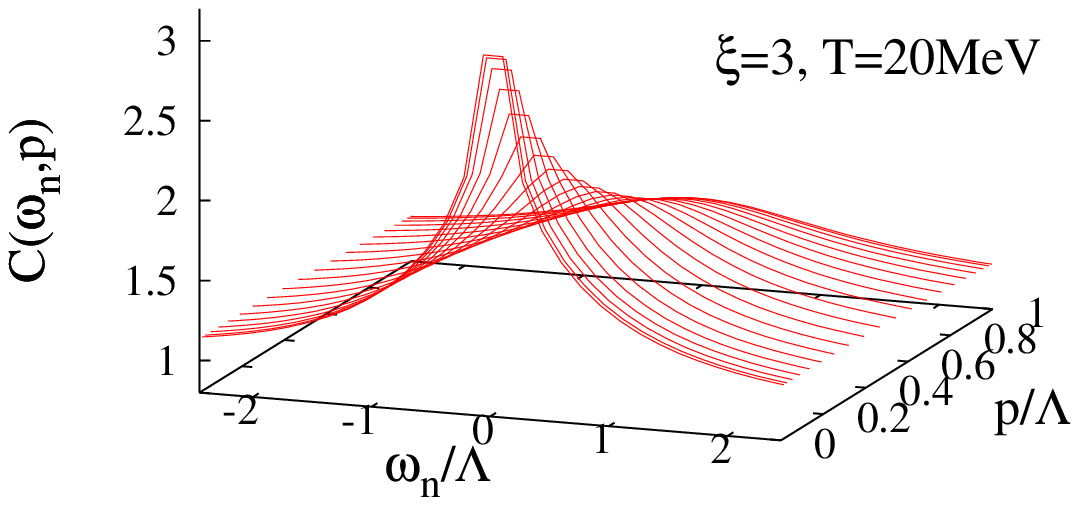}
  \includegraphics[width=5.4cm,keepaspectratio]{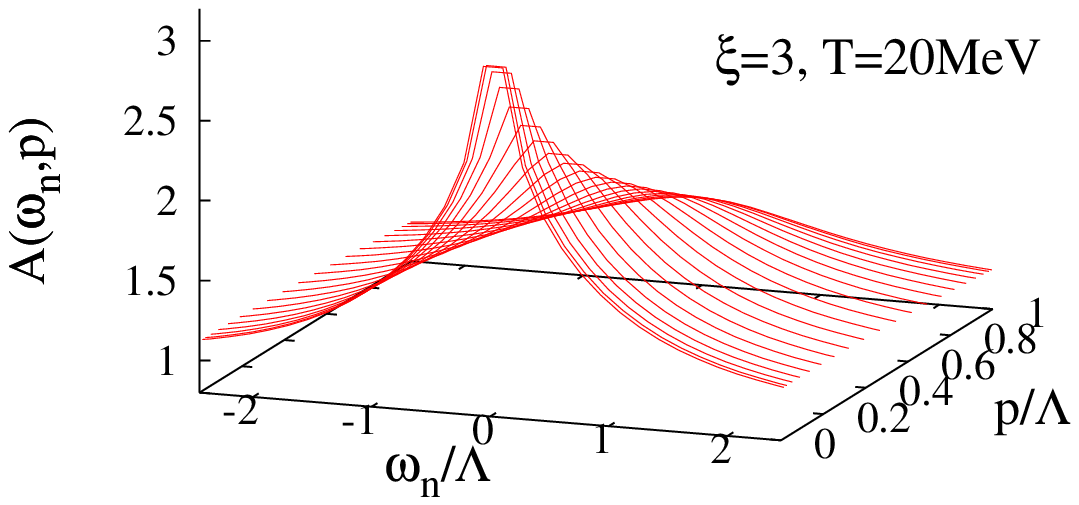}
  \includegraphics[width=5.4cm,keepaspectratio]{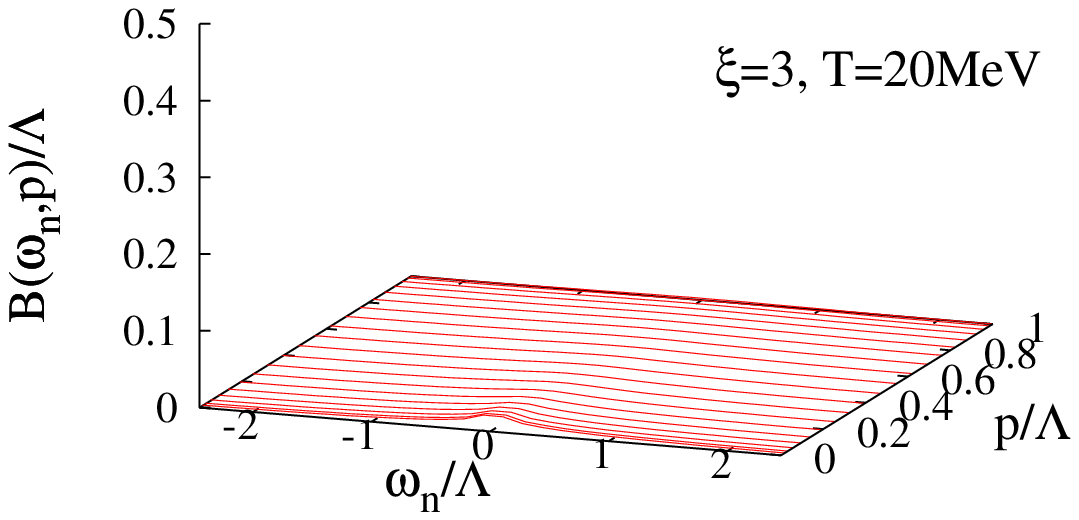}
  \includegraphics[width=5.4cm,keepaspectratio]{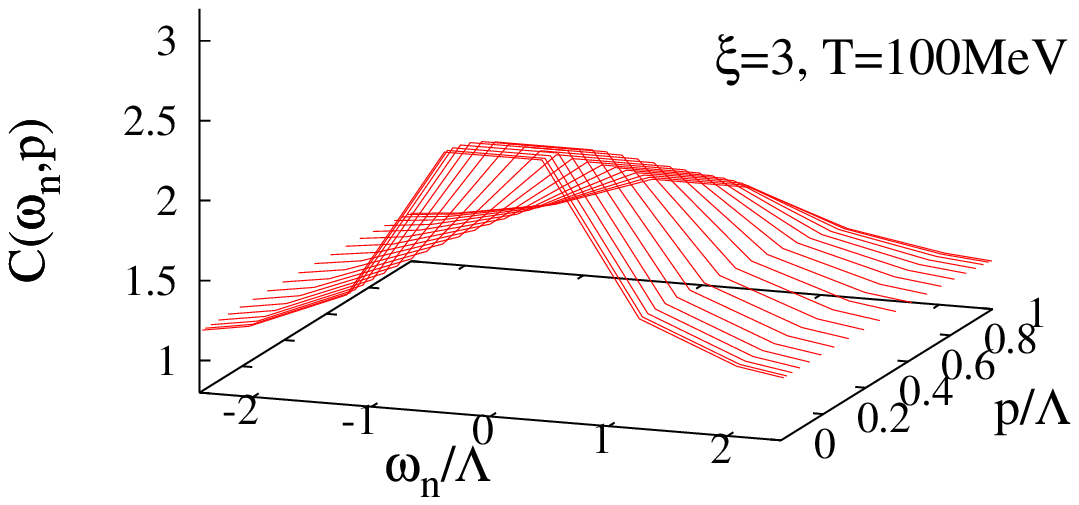}
  \includegraphics[width=5.4cm,keepaspectratio]{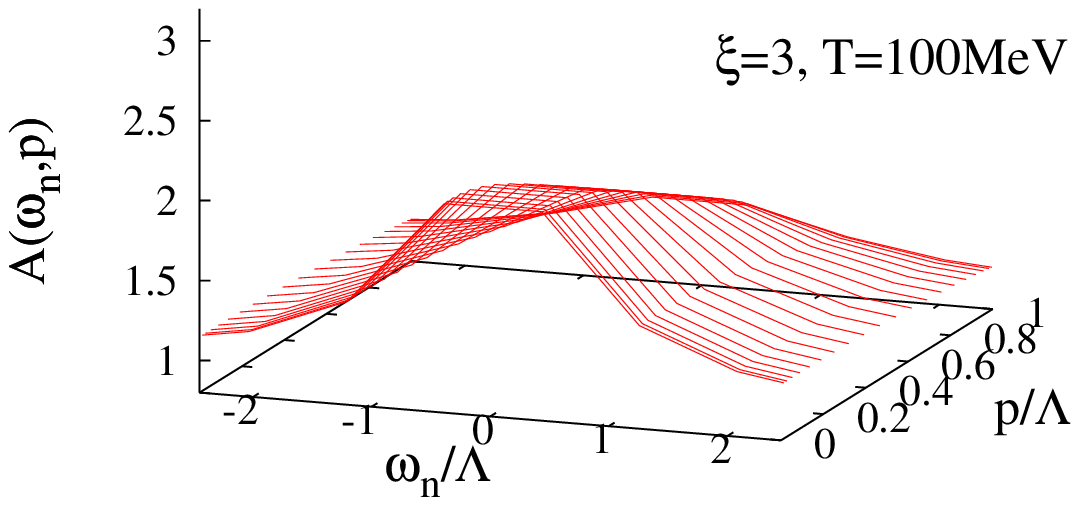}
  \includegraphics[width=5.4cm,keepaspectratio]{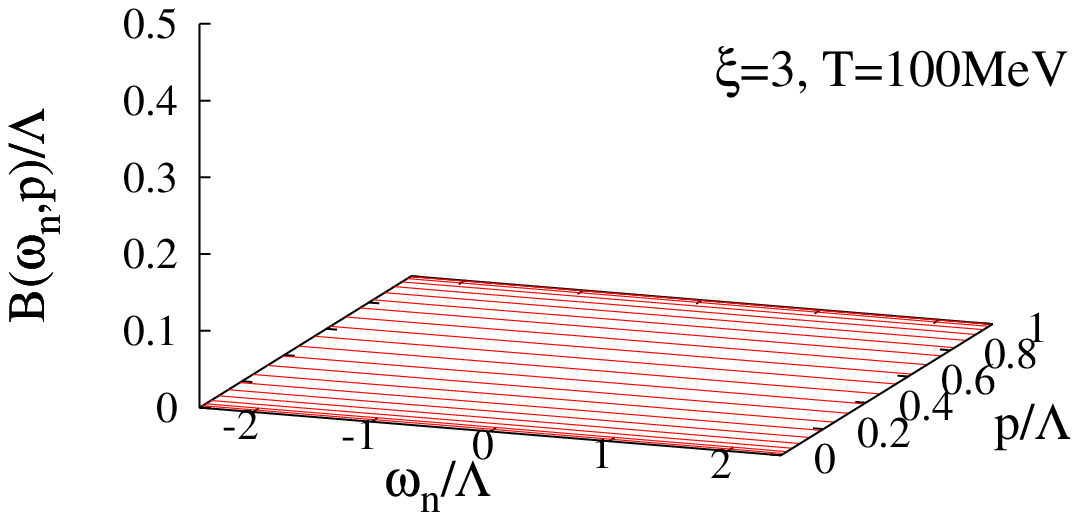}
  \caption{\label{fig:ABC}
  Numerical solutions of $C, A, B$ for
  $\alpha=2$,  $\xi=0$, $1$, $3$ at $T=20$ and $100$MeV.}
\end{center}
\end{figure}

%%%%%%%%%%%%%%%%%%%%%%%%%%%%%%
Figure~\ref{fig:ABC} shows the numerical solutions of the SDE
for $\alpha =2$ and various gauge at low and relatively high
temperature, $T=20$ and $100$MeV.
%%%%%%
%%%%%%
One notes that the values for $C$ and $A$ are almost $1$ for the
Landau gauge, and it becomes large when $\xi$ has finite values,
$\xi=1$ and $3$. On the other hand, $B$ has the opposite tendency;
it is large for $\xi=0$ and it decreases with increasing $\xi$. This
numerically comes from the fact that the denominator of the
propagator enhances when $C$ and $A$ are large, then $B$ is small
for that case consequently. As for the temperature dependence, $B$
decreases with respect to $T$, while $C$ and $A$ do not show
drastic difference between low and high temperature. The results
for $B$ is easily understood, since the restoration of the chiral
symmetry breaking would be expected. On the other hand, the
numerical solutions for $C$ and $A$ show that they do not decrease
significantly up to intermediate temperature. We will discuss in more
detail on the temperature dependence on $C$, $A$ and $B$ in the
next subsection.

%%%%%%%%%%%%%%%%%%%%%%%%%%%%%%
\subsection{\label{subsec:gds}
Temperature dependence on the solutions}
\begin{figure}[!h]
\begin{center}
  \includegraphics[width=7.0cm,keepaspectratio]{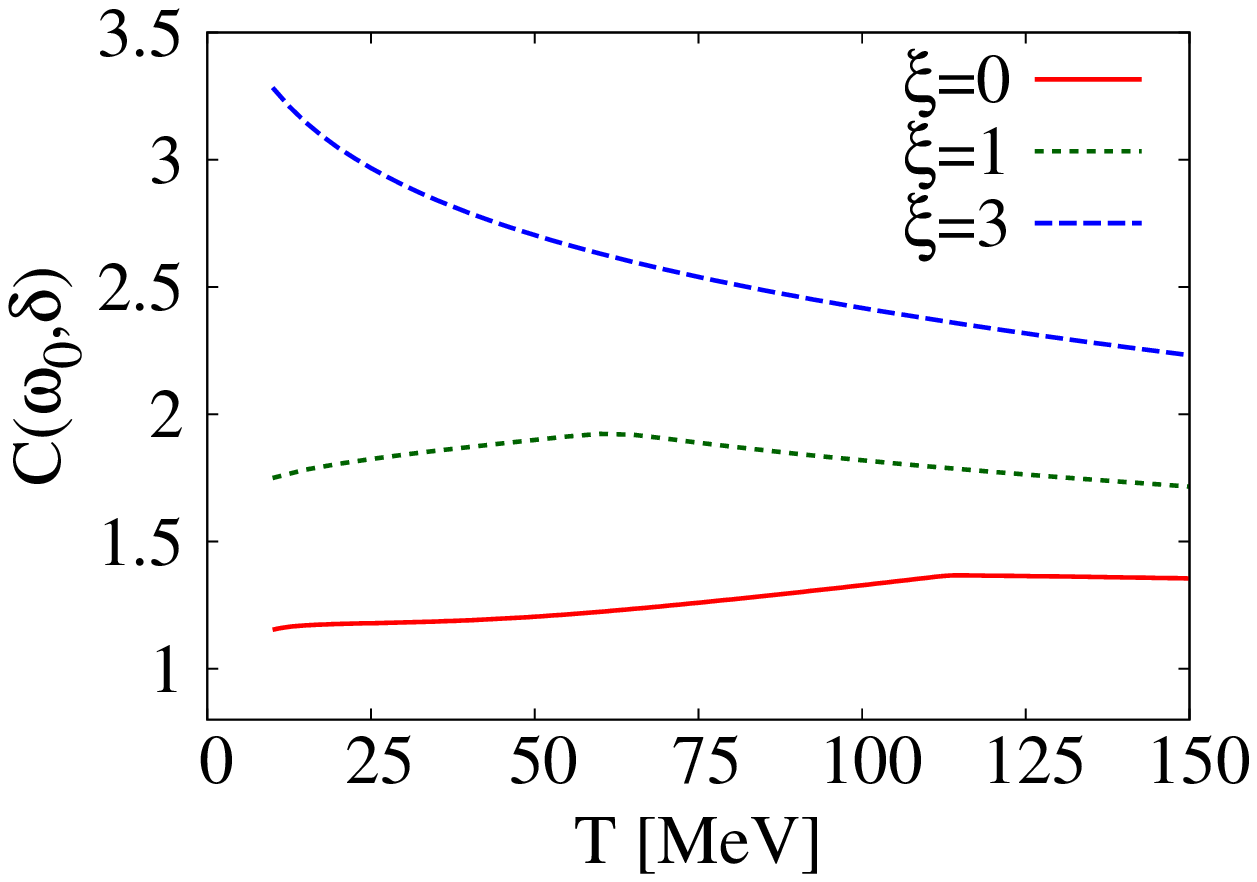}
  \includegraphics[width=7.0cm,keepaspectratio]{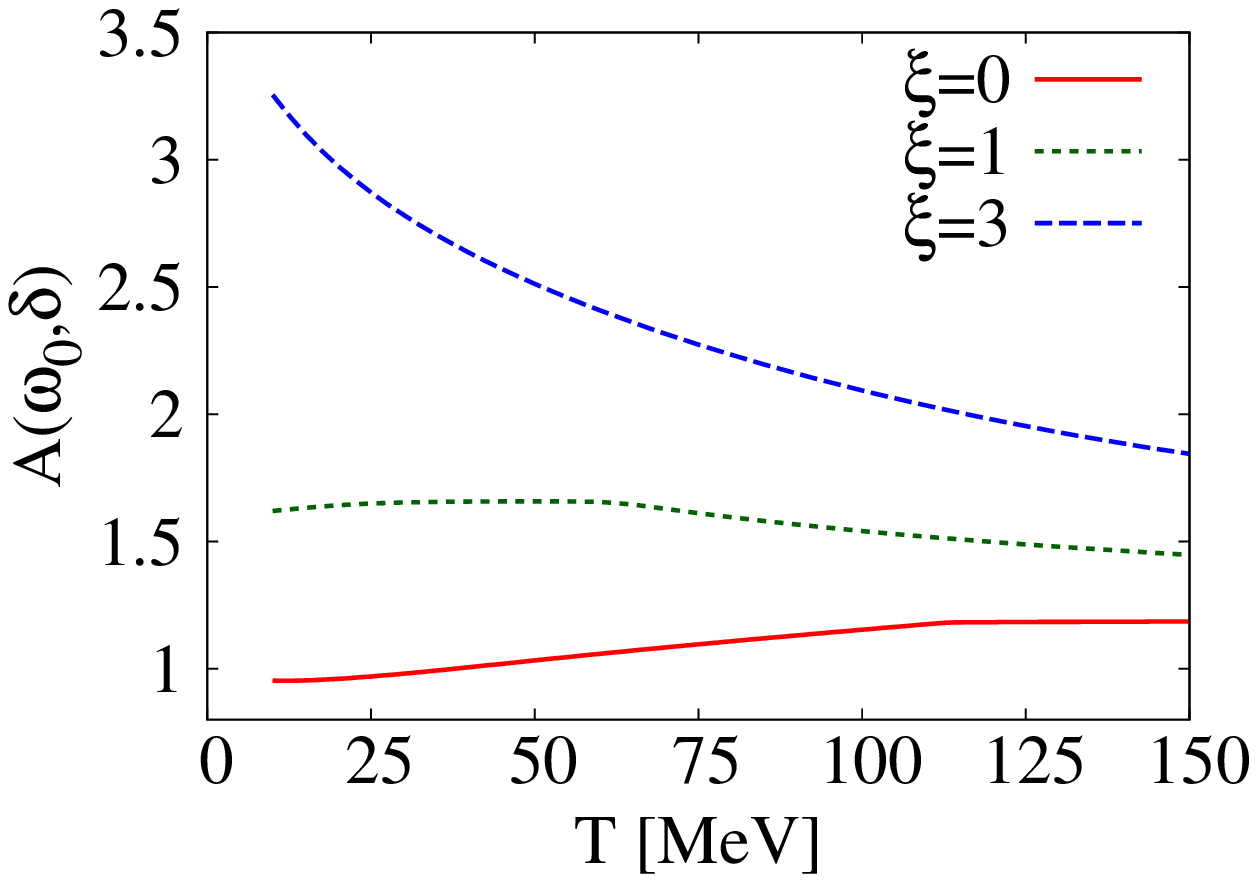}
  \includegraphics[width=7.0cm,keepaspectratio]{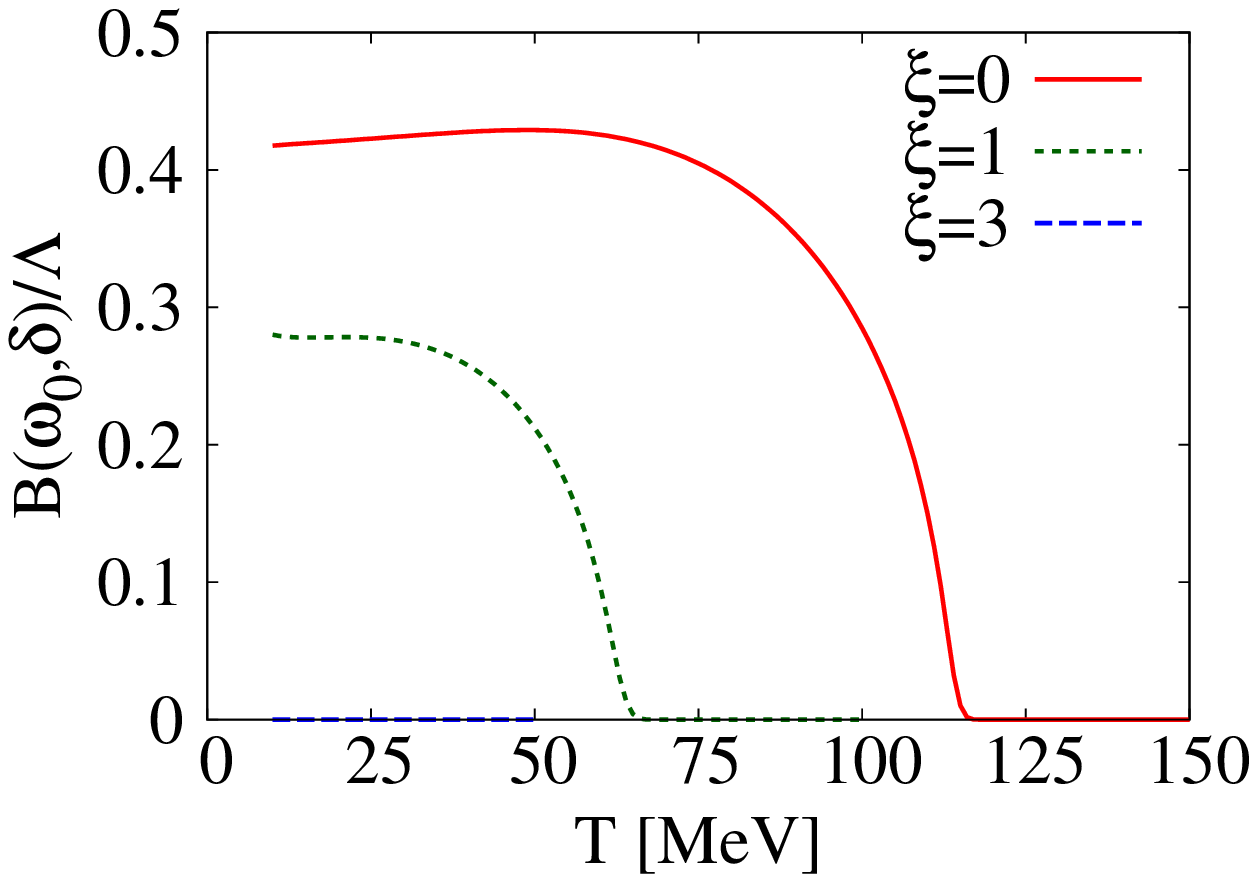}
  \includegraphics[width=7.0cm,keepaspectratio]{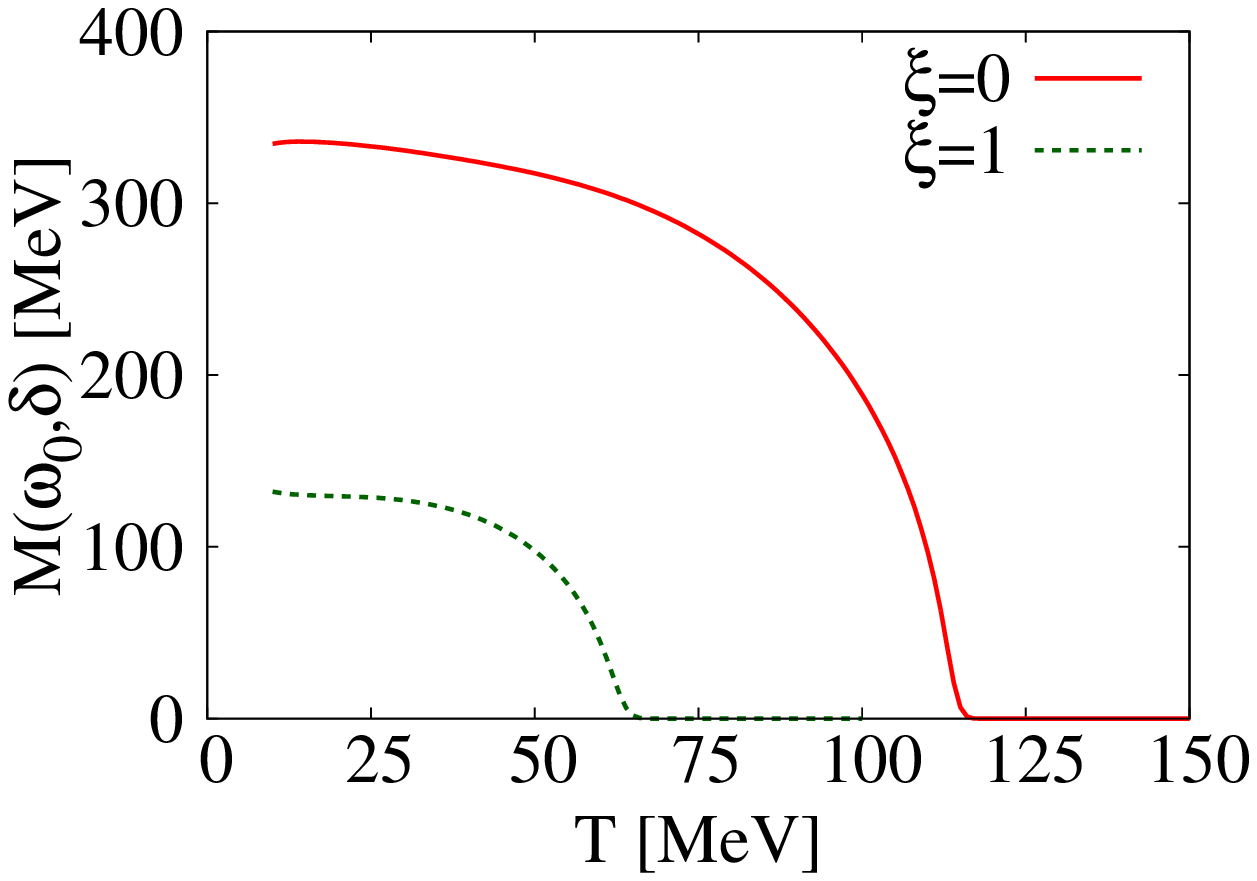}
  \caption{\label{fig:ABCMt}
  $C$, $A$, $B$ and $M$ with $\alpha=2$ for $\xi=0$, $1$ and $3$.}
\end{center}
\end{figure}

%%%%%%%%%%%%%%%%%%%%%%%%%%%%%%
Let us display the two dimensional figures so that one can see
the temperature dependence in more clearly. Figure~\ref{fig:ABCMt}
shows the results for $C(\omega_0,\delta)$, $A(\omega_0,\delta)$,
$B(\omega_0,\delta)$ and the dynamical mass $M(\omega_n,p)\equiv 
B(\omega_n,p)/A(\omega_n,p)$ as the function of the temperature.
%%%%%%
%%%%%%
We see the rapid decrease on the results for $B$ and $M$ with
increasing temperature as already mentioned in the previous
subsection, while $C$ and $A$ exhibit rather mild change;
they monotonically decreases according to $T$ for $\xi=3$, and
stay almost the same values for $\xi =0$ and $1$. The difference
between $B$ and $C$, $A$ comes from the different form of the
equations; $B$ has the form of $B = \int \md^4 q f_B(q) B(q)$ for
massless case, while $C$ and $A$ have the complex form as
$C = 1 + \int \md^4 q [f_C(q) C(q) + f_A(q) A(q)]$. Therefore,
consequently, $C$ and $A$ become close to $1$ for the case with
small values of $\xi$, which is numerically confirmed in the
above figure.

%%%%%%%%%%%%%%%%%%%%%%%%%%%%%%
\subsection{\label{subsec:critical}
Critical behavior}
%%%%%%%%%%%%%%%%%%%%%%%%%%%%%%
Finally, we are going to study the critical temperature of the chiral
phase transition through seeing the results of the dynamical mass.
For our purpose, observing the value of $M(\equiv B/A)$ is enough
since the chiral condensate is defined by,
%%%%%%
\begin{align}
  \phi = \langle \bar{\psi} \psi \rangle
  = -T \sum_{m=-\infty}^{\infty}
     \int \frac{\md^3 q}{(2\pi)^3} {\rm tr}
     \left[
         \frac{B(\omega_m, q) }{ C^2(\omega_m, q) \omega_m^2  
                 + A^2(\omega_m, q) q^2  + B^2(\omega_m, q)}
     \right],
\end{align}
%%%%%%
then it becomes zero when $M=0$ while it has non-zero value
for $M \neq 0$.

\begin{figure}[!h]
\begin{center}
  \includegraphics[width=7.0cm,keepaspectratio]{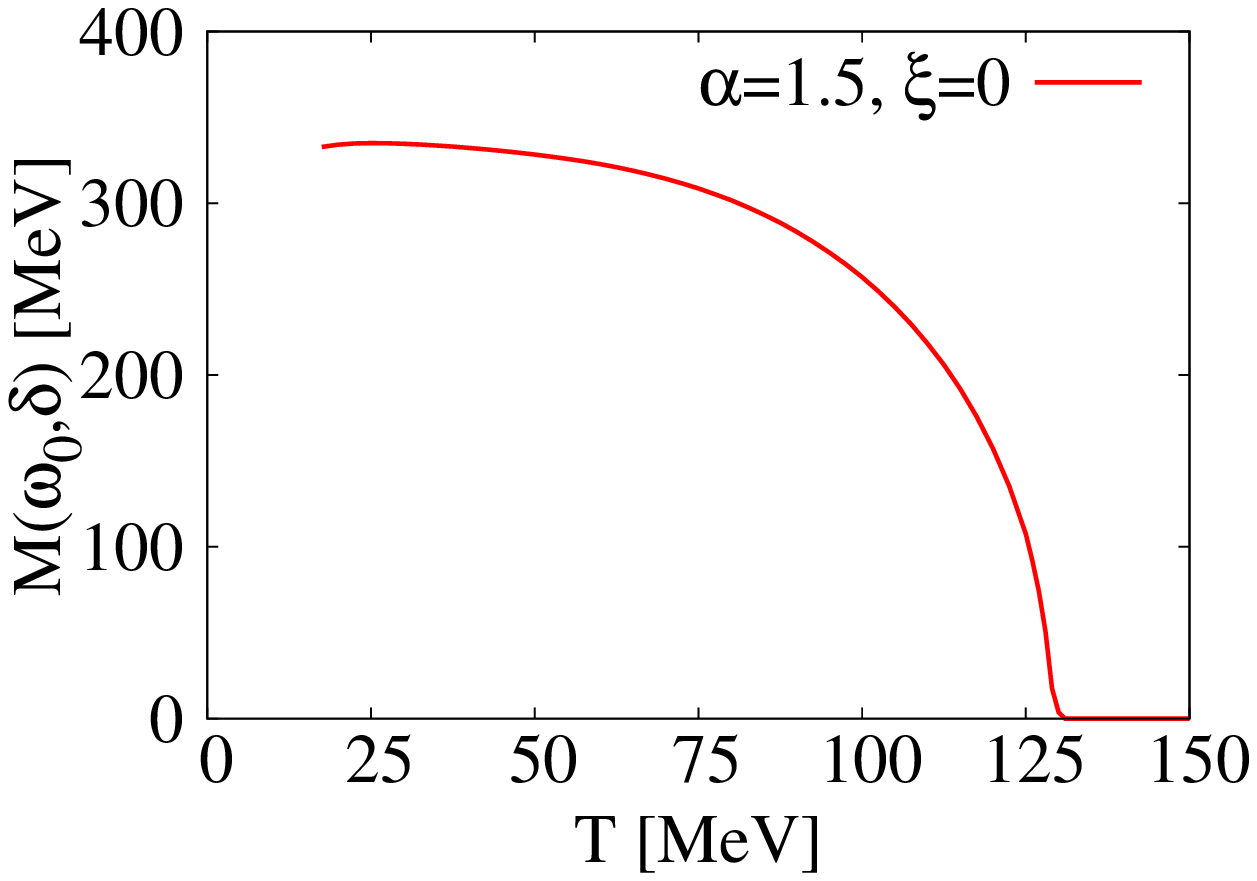}
  \includegraphics[width=7.0cm,keepaspectratio]{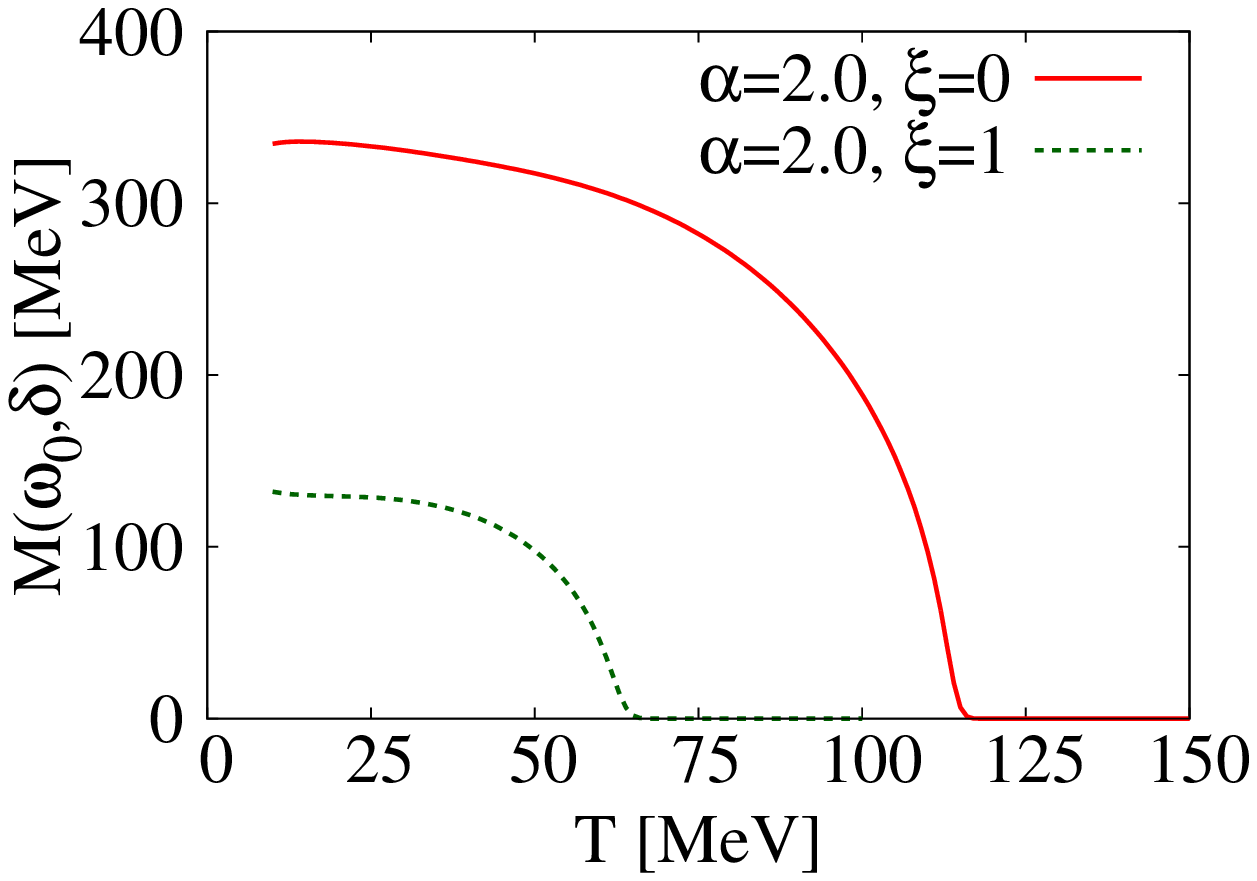}
  \includegraphics[width=7.0cm,keepaspectratio]{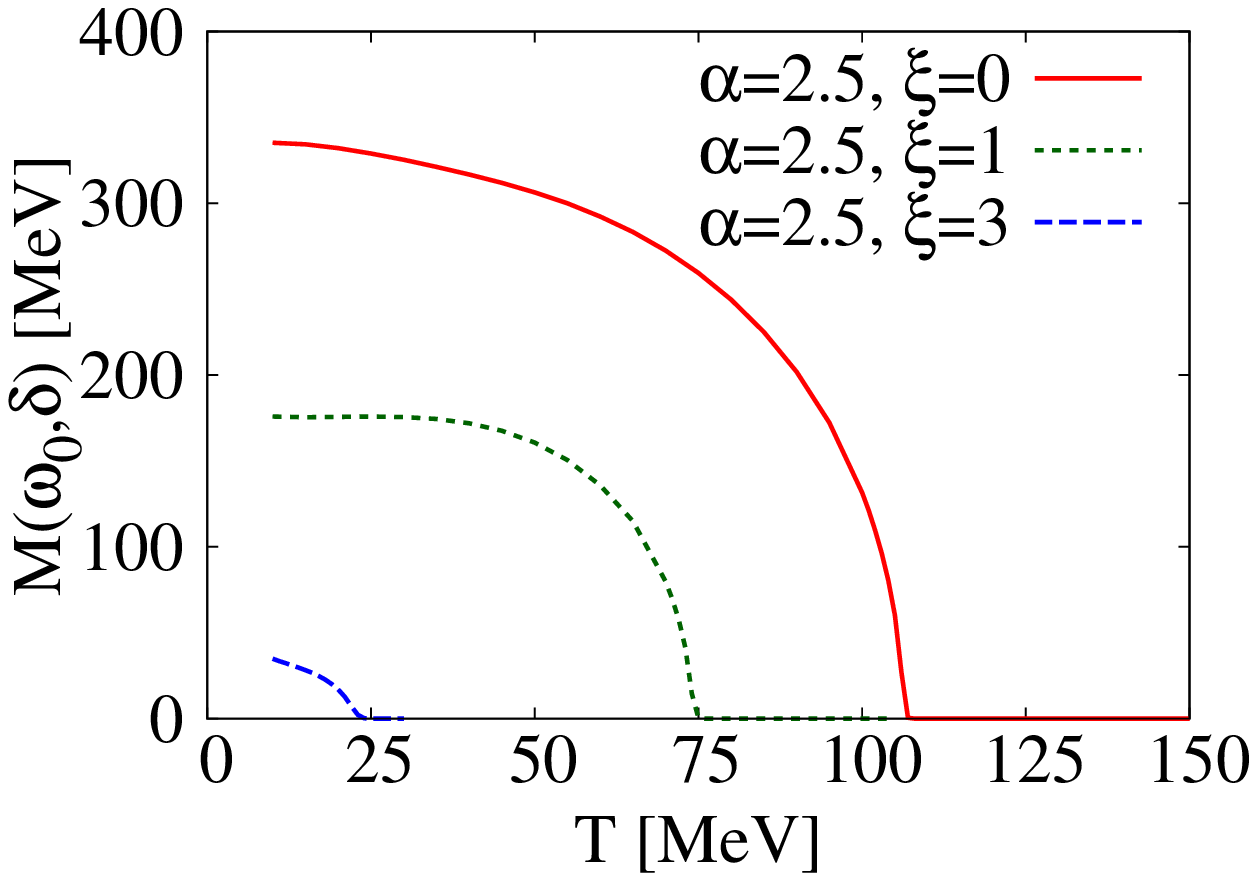}
  \includegraphics[width=7.0cm,keepaspectratio]{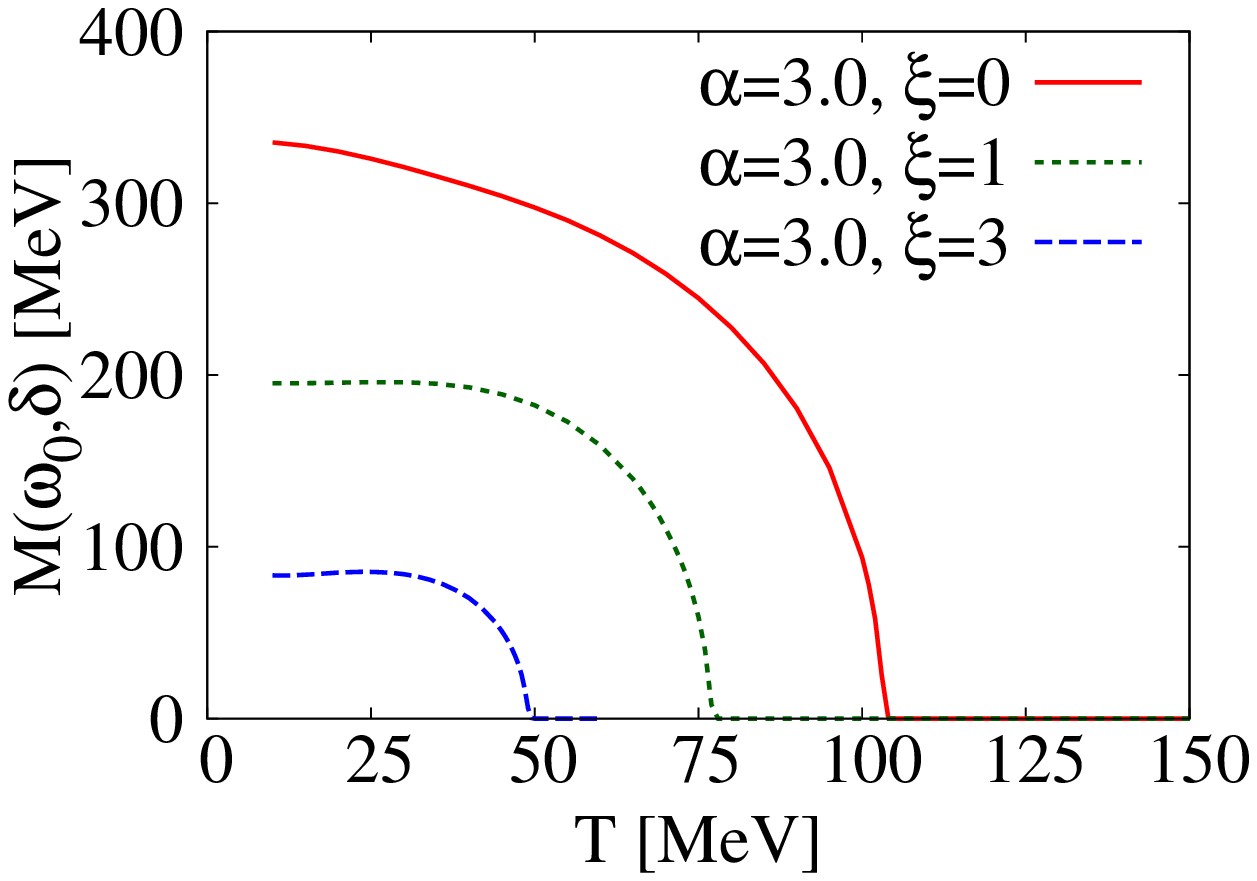}
  \caption{\label{fig:Mt}
  $M$ for various $\alpha(=1.5,\, 2,\, 2.5,\, 3)$ and $\xi(=0,\, 1,\, 3)$.}
\end{center}
\end{figure}

Figure~\ref{fig:Mt} shows the numerical results of the dynamical mass
for various couplings and gauges.
%%%%%%
%%%%%%
It should be noted that the value $M=335$MeV at low $T$ for $\xi=0$
is due to our choice of parameter, and the remaining results are
the predictions. One sees that the critical temperature, $T_c$,
does not alter drastically with changing $\alpha$ for $\xi=0$; $T_c$ is
in the range of $110 - 130$MeV for the Landau gauge.
On the other hand, $T_c$ are considerably different for other
gauges. The dynamical mass itself is not generated
for smaller coupling; $M=0$ always holds for $\alpha=1.5$ with
$\xi=1$ and $3$, and for $\alpha=2$ with $\xi=3$. From the
figure we confirm the tendency that the critical temperature
becomes small when $\xi$ is large.

To make the clearer comparison, we align the critical temperature
in Tab.~\ref{tab:Tc}.
%%%%%%%%%%%%%%%%%%%%%%%%%%
\begin{table}[h!]
\caption{Critical temperature [MeV].}
\label{tab:Tc}
\begin{center}
\begin{tabular*}{10cm}{@{\extracolsep{\fill}}ccccc}
\hline
   & $\xi=0$ & $\xi=1$ & $\xi=2$  & $\xi=3$ \\
\hline
$\alpha=1.5$ & $131$ & ---   & ---   & --- \\
$\alpha=2.0$ & $117$ & $66$ & ---   & ---  \\
$\alpha=2.5$ & $107$ & $75$ & $51$  & $24$  \\
$\alpha=3.0$ & $105$ & $78$ & $62$  & $50$ \\
\hline
\end{tabular*}
\end{center}
\end{table}
%%%%%%%%%%%%%%%%%%%%%%%%%%
We notice that the decrease of the critical temperature with respect to
$\xi$ becomes rather mild for larger $\alpha$, there the broken phase
remains for all the case as shown in the table. It may be worth mentioning
that the results for the Landau gauge exhibit similar values obtained by
the analyses in~\cite{Blank:2010bz}. The obtained values are
smaller than the expected number of $T_c \simeq 175$MeV from other
theoretical predictions, such as the lattice QCD simulations.
This is mainly due to the setting of the massless equations, and the
parameter choice based on the dynamical mass $M=335$MeV at $T=0$
in the Landau gauge.

%%%%%%%%%%%%%%%%%%%%%%%%%%%%%%
\section{\label{sec:conclusion}
Concluding remarks}
%%%%%%%%%%%%%%%%%%%%%%%%%%%%%%
We performed the systematical numerical analyses on the quenched
SDE of QCD without applying any approximation in general gauge, then
studied the gauge dependence on the critical temperature of the chiral
phase transition in this paper. We found that the critical temperature
crucially  depends on the gauge, and the parameters of the equations.
Our numerical results show smaller values on the critical temperature,
which is around $T_c \simeq 120$MeV or less, than the expected value
$T_c \simeq 175$MeV. Then we think that careful considerations on
both the gauge and parameter dependence are important when one
studies the chiral phase transition by using the SDE.

The physical predictions from the SDE should be gauge, regularization
and parameter independent in principle as mentioned in the introduction.
However, the practical equations with several approximations lead
gauge and parameter dependence as confirmed in this letter. For
studying the practical usage of the SDE, we employed the quenched
form without applying any further approximations, since it is the
simplest and frequently used equations. Concerning on the above
gauge dependence of the SDE, a lot of efforts have been made to
obtain the gauge independent
solutions (see, e.g.,~\cite{%
Rebhan:1992ak,
Kizilersu:2014ela,
Hoshino:2014nga%
}). We think this will be the future direction on the current approach
beyond the quenched level.

%%%%%%%%%%%%%%%%%%%%%%%%%%%%%%
\begin{acknowledgments}
The author thanks to T. Inagaki and M. Kohda for discussions.
The author is supported by Ministry of Science and Technology
(Taiwan, ROC), through Grant No. MOST 103-2811-M-002-087.
\end{acknowledgments}
%%%%%%%%%%%%%%%%%%%%%%%%%%%%%%


%%%%%%%%%%%%%%%%%%%%%%%%%%%%%%
\end{document}